\documentclass[aps,pra,secnumarabic,
twocolumn, groupedaddress,showpacs,amsmath,amssymb,superscriptaddress]{revtex4}

\usepackage{graphicx}
\usepackage{bm}

\begin{document}

\title{Underbarrier nucleation kinetics in a metastable quantum liquid near the spinodal }

\author{S.N. Burmistrov}
\email[]{burmi@kurm.polyn.kiae.su} \affiliation{Kurchatov Institute, 123182 Moscow, Russia} \affiliation{Tokyo
Institute of Technology, 2-12-1, O-okayama, Meguro, Tokyo 152-8551, Japan}
\author{L.B. Dubovskii}
\affiliation{Kurchatov Institute, 123182 Moscow, Russia}
\author{ Y. Okuda}
\affiliation{Tokyo Institute of Technology, 2-12-1, O-okayama, Meguro, Tokyo 152-8551, Japan}


\begin{abstract}
We  develop a theory in order to describe the effect of relaxation in a condensed medium upon the quantum
decay of a metastable liquid  near the spinodal  at low temperatures. We find that both the regime and the
rate of quantum nucleation strongly depend on the relaxation time and its temperature behavior. The quantum
nucleation rate slows down with the decrease of the relaxation time. We also discuss the low temperature
experiments on cavitation in normal $^3$He and superfluid $^4$He at negative pressures. It is the sharp
distinctions in the high frequency sound mode and in the temperature behavior of the relaxation time that make
the quantum cavitation kinetics in $^3$He and $^4$He completely different in kind.
\end{abstract}


\maketitle

\section{Introduction}
The phenomenon of macroscopic quantum nucleation  in a metastable condensed medium, which is one of possible
examples of the macroscopic quantum tunneling phenomena, has attracted a noticeable attention of researchers
during recent years \cite{r1,r2}. The range of the metastable systems in which the macroscopic quantum
nucleation is currently investigated is rather wide. These are from the helium systems involving, in
particular, solidification from the overpressurized liquid phase \cite{r3}, various types of sound-induced
nucleation \cite{r4,r4a}, phase separation of supersaturated $^3$He-$^4$He superfluid mixtures \cite{r5}, and
cavitation of bubbles at negative pressures \cite{r6,r7} to the collapse  of the Bose-Einstein condensate in a
Bose gas with attraction \cite{r8} and the formation of a quark matter in the core of neutron stars \cite{r9}.
\par
The modern representation on the decay of a meta\-stable condensed medium is associated, first of all, with
the nucleation of a seed of the stable phase and with the concept of a critical nucleus sufficient to overcome
some potential barrier due to either thermal or quantum fluctuations and then to convert the whole metastable
phase into the stable one. The decay rate due to its exponential behavior as a function of the barrier height
depends drastically on the imbalance between the phases. As a rule, it is believed that the experimentally
observable decay rates should be connected with the vicinity of a metastable medium to some instability
resulting in the corresponding reduction of a potential barrier which prevents from the formation of a
critical nucleus. In particular, in the case of the homogeneous cavitation of gas bubbles in a liquid such
instability can be associated with approaching to the spinodal line for the liquid-vapor phase transition. The
latter is one of the fundamental conclusions in the recent experiments on the quantum cavitation of bubbles in
liquid $^3$He and $^4$He at negative pressures \cite{r10}. The next important aspect of the experimental
observations is a crossover in the temperature behavior of the attainable cavitation pressure. This is treated
as an evidence for the thermal-quantum crossover in the cavitation kinetics.
\par
Unfortunately, so far the quantum decay of a meta\-stable condensed medium  close to instability has not
received an appropriate attention with the exception of works \cite{r11,r12} in which any relaxation processes
in a medium are completely ignored in consideration. However, nobody can deny an importance of involving
relaxation processes especially for the case of quantum liquids due to strong effect of temperature on the
relaxation time in quantum media at low temperatures. Here we eliminate the shortages of the previous theory
and involve the effect of relaxation processes accompanied by irreversible energy dissipation upon the
underbarrier nucleation kinetics.

\section{Energy and dispersion spectrum of density fluctuations}

\par It is long known that the critical nuclei responsible for the decay of a metastable condensed medium near
the spinodal are characterized by the relatively smooth boundaries and more extended sizes compared with the
case of nucleation near the binodal corresponding to thermodynamic equilibrium between the phases. That is
why, the van der Waals approach or gradient decomposition is usually employed for the description of
fluctuations in a medium. Accordingly, the energy of a reversible fluctuation in a liquid reads as \cite{r13,
r11}
\begin{eqnarray}\label{f1}
U[\delta\rho ]  & = & \int d^{3}r \left[\frac{\varepsilon ''(\rho )}{2}(\delta\rho )^2 + \frac{\varepsilon
'''(\rho )}{6}(\delta\rho )^3\right.\nonumber
\\
&+&\left. \frac{\lambda (\rho )}{2}(\nabla\delta\rho )^2 +\ldots \right]
\end{eqnarray}
where $\rho$ is the initial density of the homogeneous meta\-stable medium close to the density $\rho _c$ at
the spinodal and $\delta\rho = \delta\rho (\bm{r}, t)$ is the fluctuative deviation of the density. The second
derivative of the energy per unit volume of a liquid can be expressed via either compressibility $(\rho
^2\epsilon ''(\rho ) )^{-1}$ or sound velocity $c_0(\rho )$
\begin{equation}\nonumber
\epsilon ''(\rho) = c_{0}^{2}(\rho )/\rho
\end{equation}
\par The spinodal, associated with the violation of thermodynamic inequality $c_0^2(\rho ) > 0$ and
absolute instability against longwave fluctuations \cite{r14}, is determined by vanishing the sound velocity,
i.e.,
\begin{equation}\label{f3}
c_0(\rho _c) = 0
\end{equation}
The parameter $\lambda$, depending in general on density, determines a scale of the energy of inhomogeneity
and can be interpreted in terms of dispersion of the sound spectrum as a function of wave vector. Though the
energy of inhomogeneity also contributes into the interfacial tension usually measured at the binodal, one
should treat this very carefully. In fact, in this case one may need to involve next orders into the gradient
expansion since the thickness of the interface at the binodal does not much exceed several interatomic
distances and the change in density across it is noticeable with the exception of the immediate vicinity of
the liquid-vapor critical point.
\par
Involving condition \eqref{f3}, we can represent the potential energy of a density fluctuation \eqref{f1} as
\begin{eqnarray}\label{f4}
U[\delta\rho ] & = & \int d^{3}r \left[\frac{c_0^2(\rho )}{2\rho }\left( \delta\rho ^2 +
\frac{1}{3}\frac{\delta\rho ^3}{\rho -\rho _c} \right)\right. \nonumber
\\
& + & \left. \frac{\lambda }{2}(\nabla\delta\rho )^2 +\ldots\right]
\end{eqnarray}
The restriction with cubic terms in the expansion implies that the sound velocity vanishes as $c(\rho )\propto
(\rho -\rho _c)^{1/2}$ or $c(P)\propto (P-P_c)^{1/4}$ as a function of pressure in the vicinity of the
spinodal pressure $P_c$ corresponding to density $\rho _c$. At present, the genuine exponent in the behavior
of the sound velocity in liquid helium near the spinodal is unknown. Also, it is frequently proposed that
$c(P)\propto (P-P_c)^{1/3}$ and $c(\rho )\propto (\rho -\rho _c)$. This estimate is based mainly on
extrapolating the sound velocity data from the range of positive pressures to negative ones \cite{r14a}. An
explanation of such wholly satisfactory extrapolation beyond the close vicinity of the spinodal point can be
found, e.g., in \cite{r14b}. To describe the latter behavior which is more typical for the vicinity of the
liquid-vapor critical point, we must retain the terms of expansion to fourth order in $\delta\rho$. Though
this case requires a special treatment, we believe that the qualitative picture of nucleation remains faithful
with the exception of numerical coefficients in the final expressions.
\par
Provided we are interested only in thermal fluctuations, Eq.~\eqref{f1} or Eq.~\eqref{f4} is, in principle,
sufficient to determine the decay rate within the exponential accuracy since the nucleation process is mainly
governed by the saddle point of the functional of potential energy. In order to investigate the underbarrier
nucleation kinetics and to calculate the rate of quantum decay, we employ the formalism in terms of
imaginary-time path integrals and based on the use of the finite-action solutions (instantons) of equations of
motion continued to the imaginary time. For review, see, e.g., Refs.~\cite{r1,r2}. This approach of the
effective Euclidean action was used for describing quantum decay of a metastable condensed medium near the
binodal in order to incorporate the energy dissipation effect on the quantum kinetics of first-order phase
transitions at low temperatures \cite{r15,r16}.
\par
Any fluctuation as a perturbation violates the thermodynamic equilibrium in the liquid, triggering the
internal processes to recover the equilibrium. Small oscillations of the density represent a superposition of
sound waves. The character of sound processes depends strongly on a ratio between the typical inverse
frequency of density fluctuations and the typical time of relaxation processes in a liquid. The finiteness  of
the relaxation times results, in particular,  in the frequency dispersion of the sound velocity and in the
sound attenuation. Obviously, this effect is important for quantum liquids in which the relaxation times
depend significantly on temperature.
\par
In general, deriving the exact equation of sound dispersion for the whole range of frequencies is practically
unsolvable problem. Usually, in order to describe the experimental observations, the so-called
$\tau$-approximation of a single relaxation time is employed. Then, the dispersion equation, i.e., relation
between wave vector $k$ and frequency $\omega$, reads, e.g., \cite{r17}
\begin{equation}\label{f5}
c_0^2 k^2 = \omega ^2\frac{1-\imath\omega\tau}{1-\imath\omega\tau (c_{\infty}/c_0)^2}
\end{equation}
Here $\tau =\tau (T)$ is a relaxation time depending, in general, on temperature $T$. The low frequency
$\omega\tau\ll 1$ limit corresponds fully to the usual hydrodynamic sound with attenuation coefficient
\begin{equation}\nonumber
\gamma(\omega ) =\frac{\omega ^2\tau}{2c_0}\left( \frac{c_{\infty}^2}{c_0^2}-1\right)
\end{equation}
which can be expressed in terms of viscosity $\eta = (3/4)\rho c_0^2\tau (c_{\infty}^2/c_0^2 -1)$. The sound
velocity $c_{\infty}>c_0$ stands for the velocity of high frequency $\omega\tau\gg 1$ collisionless sound with
the attenuation coefficient proportional to $1/\tau$. In liquid $^3$He the high frequency limit can be
associated with the zero-sound mode and Eq.~\eqref{f5} approximates well both the frequency dispersion of
sound velocity and the attenuation within the accuracy of several percent.

\section{Quantum description and effective action}

\par
Let us return to the underbarrier nucleation. The probability of the quantum decay will be proportional to
\begin{equation}
W \propto \exp (-S/\hbar)\nonumber
\end{equation}
where $S$ is the effective Euclidean action taken for the optimum path from the entrance point under the
potential barrier to the point at which the optimum fluctuation escapes from the barrier. So, for calculating
quantum probability, we must construct the effective action determined in the imaginary time so that the
energy \eqref{f4} of a fluctuation would play a role of a potential energy. In addition, in accordance with
the principle of analytic continuation into the real time the extremum path obtained  by varying the effective
action and analytically continued from the imaginary Matsubara frequencies to real ones ($\mid\omega
_n\mid\rightarrow -\imath\omega$) should reproduce the classical equation of motion, i.e., dispersion
Eq.\eqref{f5} which the dynamics of small density fluctuations obeys in the real time.
\par
Due to small variations of the density for fluctuations near the spinodal we can describe the motion of a
liquid with the aid of the field of displacement $\bm{u}(\bm{r},t)$. In other words, we will treat the liquid
phase as an elastic medium with zero shear modulus. A relative variation of the density is related to the
change of the bulk element due to deformation
\begin{equation}\nonumber
\frac{\delta\rho}{\rho} = \frac{\rho ' -\rho}{\rho} = -\,\frac{dV' - dV}{dV'}
\end{equation}
The bulk elements after and before deformation are connected with the Jacobian according to \cite{r18}
\begin{equation}\nonumber
dV' = \parallel\delta _{ik}+\partial u_{i}/\partial r_{k}\parallel dV
\end{equation}
where $\bm{u}(\bm{r},t)$ is the displacement vector describing the deformation of a medium.
\par
Emphasize that we should employ the relation between density variation $\delta\rho$ and displacement
$\bm{u}(\bm{r},t)$ beyond the linear approximation since cubic terms in $\delta\rho$ are involved into the
expansion of the energy of a fluctuation. Second order in $\bm{u}(\bm{r},t)$ is sufficient for our purposes
\begin{equation}\label{f10}
\frac{\delta\rho}{\rho}\approx -\nabla\cdot \bm{u} +\frac{1}{2}\left[\left(\frac{\partial u_l}{\partial
r_l}\right) ^2 + \frac{\partial u_i}{\partial r_k}\frac{\partial u_k}{\partial r_i}\right]
\end{equation}
\par
As a result, we arrive at the following nonlocal effective action
\begin{widetext}
\begin{equation}\label{f11}
S[\bm{u}(\bm{r},t)] = \int _{-1/2T}^{1/2T}dtdt'\int d^3r\,
\frac{\rho}{2}\dot{\bm{u}}(\bm{r},t)D(t-t')\dot{\bm{u}}(\bm{r},t') + \int _{-1/2T}^{1/2T}dt\int d^3r\,
\left[\frac{c_0^2(\rho)}{2\rho}\left(\delta\rho ^2 +\frac{1}{3}\frac{\delta\rho ^3}{\rho -\rho _c}\right) +
\frac{\lambda}{2}(\nabla\delta\rho )^2\right]
\end{equation}
\end{widetext}
Here we put $\hbar =1$ and $k_B =1$. The nonlocal kernel $D(t)$ describing relaxation of a density fluctuation
is simply expressed in terms of its Fourier transform $D(\omega _n)$
\begin{widetext}
\begin{eqnarray}
D(t)=T\sum _{n} D(\omega _n)e^{-\imath\omega _nt} \,\, , \,\,\omega _n =2\pi nT\,\, , \,\, n=0,\, \pm 1,\, \pm
2, \ldots \nonumber
\\
D(\omega _n)=\frac{1+\mid\omega _n\mid\tau}{1+\mid\omega _n\mid\tau c_{\infty}^2/c_0^2} =\left\{
\begin{array}{ccc}
1-\mid\omega _n\mid\tau\frac{c_{\infty}^2 -c_0^2}{c_0^2} \, , &\tau c_{\infty}^2/c_0^2\ll \mid\omega _n\mid
^{-1} &
\\
\frac{1}{\mid\omega _n\mid\tau}\,\frac{c_0^2}{c_{\infty}^2}\, , &  \tau\ll\mid\omega _n\mid ^{-1}\ll\tau
c_{\infty}^2/c_0^2\, , & \text{if}\;\; c_{\infty}\gg c_0
\\
\frac{c_0^2}{c_{\infty}^2}\left( 1+\frac{c_{\infty}^2 -c_0^2}{c_{\infty}^2}\frac{1}{\mid\omega
_n\mid\tau}\right)\, , &  \mid\omega _n\mid ^{-1}\gg\tau &
\end{array}
\right. \label{f12}
\end{eqnarray}
\end{widetext}
The latter is in the full correspondence with Eq.~\eqref{f5}. The appearance of $\mid\omega _n\mid$ in
\eqref{f12} agrees with the general rule of analytic continuation $\omega\rightarrow \imath\mid\omega _n\mid$
for the Fourier transforms of retarded correlators in the course of transformation from the physical real time
to imaginary time. Note that the case of zero relaxation time $\tau =0$ corresponds to the action considered
in \cite{r11} within the linear approximation $\delta\rho /\rho =-\nabla\cdot\bm{u}$ in \eqref{f10}. In
\cite{r11} the nucleation dynamics was assumed to be completely reversible and thus any possible effect of
relaxation and energy dissipation upon the underbarrier nucleation was ignored.
\par
As a next step, it is convenient to introduce the dimensionless units
\begin{eqnarray}\label{f13}
\chi =\frac{1}{2}\frac{\delta\rho}{\rho -\rho _c}\, , \,\, x=\frac{r}{l}\, ,\,\, \eta=\frac{t}{t_0}\, ,
\nonumber
\\
\bm{v}=\frac{\bm{u}}{u_0}\, , \,\, T'=Tt_0\, , \,\, \Omega _n =\omega _nt_0
\end{eqnarray}
Here we put
\begin{equation}\nonumber
l^2 = \frac{\lambda\rho}{c_0^2(\rho )}\, , \,\, u_0=2l\,\frac{\rho -\rho _c}{\rho}\, , \, \mbox{and} \,\,
t_0=\frac{l}{c_0(\rho )}
\end{equation}
Accordingly, relation \eqref{f10} goes over into
\begin{equation}\label{f14}\nonumber
\chi = -\nabla\cdot \bm{v} +\frac{\rho -\rho _c}{\rho}\left[\left(\frac{\partial v_l}{\partial x_l}\right) ^2
+ \frac{\partial v_i}{\partial x_k}\frac{\partial v_k}{\partial x_i}\right]
\end{equation}
and action \eqref{f11} reduces to
\begin{equation}\label{f16}\nonumber
S=S_0s[\bm{v}(\bm{x},\eta)]
\end{equation}
where the dimensional factor is given by
\begin{equation}\label{f17}
S_0 = 4t_0l^3\frac{c_0^2(\rho )}{\rho}(\rho -\rho _c)^2 =4\lambda ^2\rho\frac{(\rho -\rho _c)^2}{c_0^3(\rho )}
\end{equation}
The dimensionless action decomposed to third order in displacement $\bm{v}$ reads
\begin{widetext}
\begin{eqnarray}\label{f18}
s[\bm{v}(\bm{x},\eta)] & = & \int _{-1/2T'}^{1/2T'}d\eta d\eta '\int d^3x\,
\frac{1}{2}\dot{\bm{v}}(\bm{x},\eta)D(\eta-\eta ')\dot{\bm{v}}(\bm{x},\eta ')
\\
& + & \int _{-1/2T'}^{1/2T'}d\eta\int d^3x\, \left\{\frac{1}{2}(\nabla(\nabla\cdot\bm{v}))^2
+\frac{1}{2}(\nabla\cdot\bm{v})^2 - \frac{1}{3}\left[(\nabla\cdot\bm{v})^3 -3\,\frac{\rho -\rho
_c}{\rho}\left( (\nabla\cdot\bm{v})^3 +\frac{\partial v_i}{\partial x_k}\frac{\partial v_k}{\partial
x_i}\frac{\partial v_l}{\partial x_l}\right)\right]\right\}\nonumber
\end{eqnarray}
\end{widetext}
where the Fourier transform of $D(\eta )$ is given by
\begin{equation}\nonumber
D(\Omega _n)=\frac{1+\mid\Omega _n\mid\tau/t_0}{1+\mid\Omega _n\mid\tau c_{\infty}^2/t_0c_0^2}
\end{equation}
The reduction of the coefficient in the front of $(\nabla\cdot\bm{v})^3$ leads to increasing the extremum
value of the action. However, keeping in mind the vicinity to the spinodal,  we will neglect below the cubic
terms proportional to $(\rho -\rho _c)/\rho \ll 1$. Note only that the involvement of these cubic terms yields
a correction of the same order of $(\rho -\rho _c)/\rho $ into the final answer and can be treated as a
perturbation.

\section{Quantum nucleation rate}

\par
In this section we will analyze  extremum paths of effective action \eqref{f18} and, correspondingly,
determine the nucleation  rate. The exact determination of extrema is a rather complicated problem. We start
from the limiting cases. First, we mention the time-independent path entailing the classical Arrhenius law for
the nucleation rate. From \eqref{f17} and \eqref{f18} one readily obtains
\begin{eqnarray}\nonumber
S_{cl} & = & U_0/T
\\
U_0 & = & 4(\lambda ^3\rho )^{1/2}\,\frac{(\rho -\rho _c)^2}{c_0(\rho )}\, f_0 \nonumber
\end{eqnarray}
where $U_0$ plays a role of the potential barrier for nucleation and $f_0$ is the extremum saddle-point value
of the functional for the potential energy given in dimensionless units
\begin{equation}\nonumber
f[\chi (\bm{x})] = \int d^3x \left( \frac{1}{2}(\nabla\chi )^2 +\frac{1}{2}\chi ^2 +\frac{1}{3}\chi ^3\right)
\end{equation}
The numerical solution for a critical fluctuation  of the spherical symmetry results in $f_0\approx$ 43.66.
(This value is about 9\% larger than those used in \cite{r11,r12,r13}.) The amplitude at the center is $\chi
_{\text{cl}}(x=0)\approx$ $-$4.19 and becomes a half as much at the distance $x\approx$ 1.22 \cite{r19}. The
spatial behavior of critical fluctuation $\chi _{\text{cl}}(x)$ is shown in Fig.~1.
\begin{figure}
\includegraphics[scale=1]{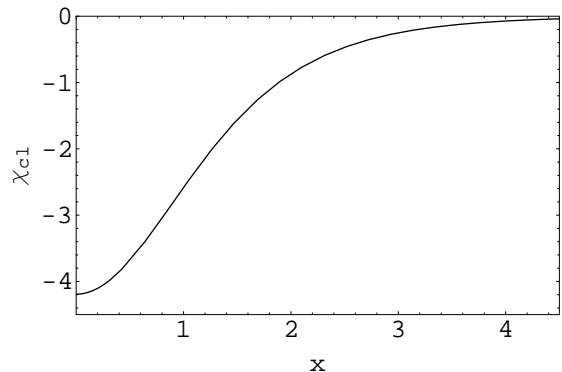}
\caption{The density profile of the thermal critical fluctuation.}
\end{figure}
\par
Let us turn now to the case of zero temperature and consider the extremum value of effective action as a
function of relaxation time $\tau$. Action $s[\bm{v}(\bm{x},\eta ]$ \eqref{f18} on the neglect of the term
proportional to $(\rho -\rho _c)/\rho$ was analyzed in \cite{r11} for the relaxation time $\tau =0$ at $T=0$.
The extremum value $s_0(0)$ for a critical quantum fluctuation was estimated as $s_0(0)\approx$ 160. The
space-time behavior of the density for the quantum critical fluctuation at zero temperature is shown in
Fig.~2.
\begin{figure}
\includegraphics[scale=1]{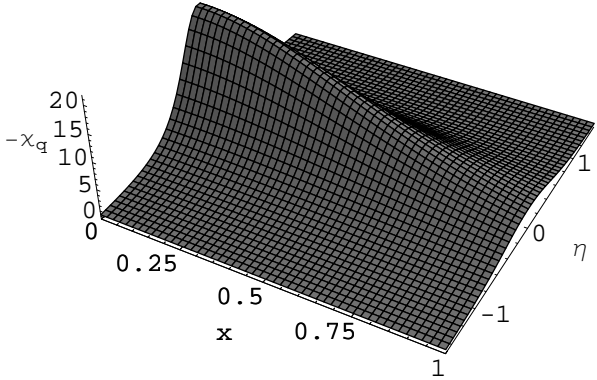}
\caption{The space-time profile of the density plotted with the minus sign, illustrating  quantum critical
fluctuation at zero temperature.}
\end{figure}
\par
The other limiting case, namely, infinite time of relaxation, can easily be reduced to the case of zero time
by redefining factor $\rho$ as $\rho _{\infty}=\rho c_0^2/c_{\infty}^2$ in the term with the kinetic energy in
\eqref{f11}. This results in the exact relation between the cases $\tau =0$ and $\tau =\infty$ since both $S$
and $T'$ are proportional to $\sqrt{\rho}$
\begin{equation}\nonumber
S_{\infty}(T) = \frac{c_0}{c_{\infty}}\, S_0\left(\frac{c_0}{c_{\infty}}T\right)
\end{equation}
 Accordingly, the relation between the extremum paths for $\tau=0$ and $\tau=\infty$ reads
\begin{equation}\label{f23}
\bm{v}_{\infty}(\bm{x},\,\eta )=\bm{v}_0(\bm{x},\,\eta c_{\infty}/c_0)
\end{equation}
\par
Employing the limiting expressions from \eqref{f12} for kernel $D(\omega _n)$ and the relation
\begin{equation}\label{f24}
T\sum _{n} |\omega _n|\text{e}^{-\imath\omega _nt} = -\,\frac{\pi T^2}{\sin ^2\pi Tt}
\end{equation}
we can estimate the effective action within first-order approximation of the theory of perturbations in two
limiting cases of small $\tau\ll t_0c_0^2/c_{\infty}^2$ and large $\tau\gg t_0c_0/c_{\infty}$ relaxation time.
For small relaxation times $\tau\rightarrow 0$, we find that correction $\Delta S_0(T)$ is negative and,
therefore, finite time of relaxation facilitates nucleation of new phase as compared with the case of zero
time of relaxation
\begin{widetext}
\begin{eqnarray}\label{f25}
S(T) & = & S_0(T)+\Delta S_0(T)
\\
\frac{\Delta S_0}{S_0} & = & -\,\frac{c_{\infty}^2-c_0^2}{\sqrt{\lambda\rho}}\, \tau\, \frac{1}{4\pi}
\int_{-1/2T'}^{1/2T'}d\eta d\eta '\int d^3x\, \left(\dot{\bm{v}}_0(\bm{x},\eta) -\dot{\bm{v}}_0(\bm{x},\eta
')\right) ^2\,\frac{(\pi T')^2}{\sin ^2\pi T'(\eta -\eta ')} \nonumber
\end{eqnarray}
\end{widetext}
where we used notation $\dot{\bm{v}}_0 =\partial\bm{v}_0/\partial\eta$. The other opposite case $\tau
^{-1}\rightarrow 0$ yields
\begin{widetext}
\begin{eqnarray}\label{f26}
S(T) & = & S_{\infty}(T)+\Delta S_{\infty}(T)
\\
\frac{\Delta S_{\infty}}{S_{\infty}} & = &
\frac{c_{\infty}^2-c_0^2}{c_{\infty}^2}\,\frac{\sqrt{\lambda\rho}}{c_0c_{\infty}}\,\frac{1}{\tau}\,
\frac{1}{4\pi} \int_{-1/2T''}^{1/2T''}d\eta d\eta '\int d^3x\, \left(\bm{v}_0(\bm{x},\eta)
-\bm{v}_0(\bm{x},\eta ')\right) ^2\,\frac{(\pi T'')^2}{\sin ^2\pi T''(\eta -\eta ')} \nonumber
\end{eqnarray}
\end{widetext}
Here we  involved Eq.~\eqref{f23} and denoted  $T''=T'c_0/c_{\infty}$. The sign of this correction is
positive.
\par
Concerning the intermediate case of relaxation time $t_0c_0^2/c_{\infty}^2\ll\tau\ll t_0c_0/c_{\infty}$ which
can be prominent only provided $c_{\infty}\gg c_0$, we should mention that the character of quantum nucleation
in this case is different in kind from the two limiting ones considered above. The point is that nonlocal
kernel $\Omega ^2D(\Omega )$ responsible for the tunneling process becomes effectively proportional to
$\mid\Omega\mid$ instead of $\Omega ^2$. Thus the dynamics of underbarrier nucleation changes from the usual
one governed by the kinetic energy term to the overdamped viscous type. The functional dependence on the
physical parameters can readily be obtained by redefining time $\eta$ from \eqref{f13} as
\begin{equation}\nonumber
\eta \rightarrow \frac{c_0^2}{c_{\infty}^2}\,\frac{t_0}{\tau}\eta =\frac{c_0^2}{c_{\infty}^2}\,\frac{t}{\tau}
\end{equation}
and involving expansion \eqref{f12} for $c_{\infty}\gg c_0$. Then we have using \eqref{f24} at $T=0$
\begin{equation}\label{f26a}
S_{\tau}(0)=S_0\,\frac{c_0^2}{c_{\infty}^2}\,\frac{t_0}{\tau}\, s_{\tau}[\bm{v}(\bm{x},\,\eta )]
\end{equation}
where the dimensionless action $s_{\tau}[\bm{v}(\bm{x},\,\eta )]$ reads
\begin{eqnarray*}
s_{\tau}[\bm{v}(\bm{x},\,\eta )]=\frac{1}{4\pi}\int_{-\infty}^{\infty}\!\!d\eta d\eta '\!\int\! d^3x
\frac{(\bm{v}(\bm{x},\eta) -\bm{v}(\bm{x},\eta ')) ^2}{(\eta -\eta ')^2}
\\
 + \int_{-\infty}^{\infty}\!\! d\eta\int\!
d^3x \left( \frac{1}{2}(\nabla\chi )^2 +\frac{1}{2}\chi ^2 +\frac{1}{3}\chi ^3\right)
\end{eqnarray*}
Thus, as the relaxation time increases, the extremum value of effective action decreases gradually and reduces
by factor $c_0/c_{\infty}$ with saturation at $\tau =\infty$. In the intermediate region
$t_0c_0^2/c_{\infty}^2\ll\tau\ll t_0c_0/c_{\infty}$ the action falls approximately as $1/\tau$.
\par
Analyzing  temperature behavior of the effective action, of course one should take into account the
temperature dependence of the thermodynamic quantities involved into the nucleation process. However, most
important aspect of the temperature behavior for the nucleation probability is associated with the temperature
dependence of relaxation time $\tau =\tau (T)$. It is obvious that one should compare the relaxation time
$\tau (T)$ with the typical time of the order of $t_0$ necessary for the underbarrier growth of a nucleus.
Involving the  frequency dispersion in \eqref{f12}, we introduce two temperatures $T_{\tau 1}$ and $T_{\tau
2}$ defined according to
\begin{eqnarray}\label{f27}
\tau (T_{\tau 1})\sim \frac{c_0}{c_{\infty}}t_0 =\frac{\sqrt{\lambda\rho}}{c_0c_{\infty}}
\\
\tau (T_{\tau 2})\sim \frac{c_0^2}{c_{\infty}^2}t_0 =\frac{\sqrt{\lambda\rho}}{c_{\infty}^2} \nonumber
\end{eqnarray}
Keeping in mind that usually the relaxation time does not diminish as the temperature lowers, we thus suppose
$T_{\tau 1}<T_{\tau 2}$. Note that these parameters enter naturally Eq.~\eqref{f25} and Eq.~\eqref{f26},
determining the order-of-magnitude correction comparable with action $S_0$ and $S_{\infty}$, respectively. The
physical meaning of temperatures $T_{\tau 1}$ and $T_{\tau 2}$ lies in separating the high frequency  type of
underbarrier nucleation processes at $T\ll T_{\tau 1}$ from the low frequency type at $T\gg T_{\tau 2}$. The
magnitude of the transient region depends on a ratio $c_{\infty}/c_0$.
\par
Let us now turn to the estimate of the thermal-quantum crossover temperature $T_q$. At first, we assume that
$T_q$ is above $ T_{\tau 2}$. Then, using $S_0(0)$, we can estimate the thermal-quantum crossover temperature
approximately as
\begin{equation}\label{f28}
T_q\sim T_{q\, 0}\approx \frac{U_0}{S_0(0)}
=\frac{f_0}{s_0}\,\frac{1}{t_0}=0.27\frac{c_0^2}{\sqrt{\lambda\rho}}\; ,\; \text{if}\;\; t_0T_{\tau 2}\ll 1
\end{equation}
In the opposite case provided $T_q$ is below $ T_{\tau 1}$ we should employ $S_{\infty}$. This yields
\begin{equation}\label{f29}
T_q\sim T_{q\, {\infty}}=\frac{c_{\infty}}{c_0}T_{q\,0}\; ,\; \text{if}\;\; \frac{c_0}{c_{\infty}}t_0T_{\tau 1
}\gg 1
\end{equation}
In the intermediate case, assuming $T_{\tau 1}<T_q<T_{\tau 2}$, we should use Eq. \eqref{f26a} to estimate the
thermal-quantum crossover temperature. Then we find
\begin{equation}\label{f33}
T_q\sim \frac{c_{\infty}^2}{c_0^2}\,\frac{\tau}{t_0}\, T_{q0}\, , \;\text{if}\;\; t_0T_{\tau 1}\ll
\frac{c_{\infty}^2}{c_0^2}\,\frac{\tau}{t_0}\ll t_0T_{\tau 2}
\end{equation}
Thus, the larger the relaxation time, the higher the crossover temperature.
\par
In order to treat the thermal-quantum crossover  more detailed, we consider the stability of the classical
path $\bm{v_{\text{cl}}(\bm{x})}$ with respect to its small perturbations $\delta\bm{v}(\bm{x},\,\eta)$,
representing an arbitrary path as $\bm{v}(\bm{x},\,\eta
)=\bm{v}_{\text{cl}}(\bm{x})+\delta\bm{v}(\bm{x},\,\eta)$ and, correspondingly, $\chi (\bm{x},\,\eta )=\chi
_{\text{cl}}(\bm{x},\,\eta )+\delta\chi (\bm{x},\,\eta )$. Then we have, being $\chi =-\nabla\cdot\bm{v}$,
\begin{widetext}
\begin{eqnarray*}
s[\delta\bm{v}(\bm{x},\,\eta )] & = & \frac{f_0}{T'}+  \int _{-1/2T'}^{1/2T'}d\eta d\eta '\int d^3x\,
\frac{1}{2}\delta\dot{\bm{v}}(\bm{x},\eta)D(\eta-\eta ')\delta\dot{\bm{v}}(\bm{x},\eta ')
\\
  & + &  \int _{-1/2T'}^{1/2T'}d\eta\int d^3x\, \left[\frac{1}{2}(\nabla\,\delta\chi )^2
+(1+2\chi _{\text{cl}}(\bm{x}))\frac{\delta\chi ^2}{2} + \frac{\delta\chi ^3}{3} \right]
\end{eqnarray*}
\end{widetext}
To find the point of instability in temperature, it is sufficient to retain quadratic terms alone. Turning to
Fourier representation in time,
\begin{equation}\nonumber
\delta\bm{v}(\bm{x},\,\eta )=T'\sum _{n}\delta\bm{v}_n(\bm{x})\text{e}^{-\imath\Omega _n\eta}\; , \;\;
\delta\bm{v}_{-n}=\delta\bm{v}_n^{*}
\end{equation}
we arrive at the expansion
\begin{widetext}
\begin{equation}\nonumber
s[\delta\bm{v}_n(\bm{x})]= \frac{f_0}{T'}+\int d^3x\,\left\{\frac{1}{2}T'\sum _{n}\mid\Omega _n\mid ^2D(\Omega
_n)\delta\bm{v}_n\delta\bm{v}_{-n} +\nabla\delta\chi _n\cdot\nabla\delta\chi _{-n} +(1+2\chi
_{\text{cl}}(\bm{x}))\delta\chi _n\delta\chi _{-n} +\ldots \right\}
\end{equation}
\end{widetext}
Next, we decompose an arbitrary perturbation $\delta\bm{v}_n(\bm{x})$ into a series
\begin{equation}\nonumber
\delta\bm{v}_n(\bm{x})=\sum _{\alpha} C_{n,\,\alpha}\bm{w}_{\alpha}(\bm{x})
\end{equation}
over a complete orthonormal set  of eigenfunctions $\bm{w}_{\alpha}$ of the equation
\begin{equation}\label{f34}
\nabla [(-\nabla ^2 +(1+2\chi _{\text{cl}}(\bm{x}))(-\nabla\cdot\bm{w})]=E\bm{w}
\end{equation}
Then one has retaining quadratic terms alone
\begin{equation}\nonumber
s[ C_{n,\,\alpha}]=\frac{f_0}{T'}  + \frac{T'}{2}\sum _{n,\,\alpha}\left\{\mid\Omega _n\mid ^2D(\Omega
_n)+E_{\alpha}\right\}\mid C_{n,\,\alpha}\mid ^2
\end{equation}
where $E_{\alpha}$ is the eigenvalues of Eq.~\eqref{f34}. Provided the expression in braces becomes negative,
at least, for one mode, the classical path becomes absolutely unstable and the crossover to the quantum path
dependent on time is unavoidable. As the temperature lowers, the mode which first becomes unstable is a mode
with $n=1$ and with $\alpha =1$ corresponding to the minimum negative value of $E_{\alpha}$. Thus the
temperature $T_1$ of absolute instability of the classical path is determined by
\begin{equation}\nonumber
\mid\Omega _1\mid ^2D(\Omega _1)=\mid\Omega _1\mid ^2\frac{1+\mid\Omega _1\mid\tau/t_0}{1+\mid\Omega
_1\mid\tau c_{\infty}^2/t_0c_0^2}=-E_1
\end{equation}
Solving equation yields
\begin{widetext}
\begin{equation}\label{f41}
T_1=\left\{
\begin{array}{lcc}
\frac{\sqrt{\mid
E_1\mid}}{2\pi}\,\frac{1}{t_0}\left(1+\frac{c_{\infty}^2-c_0^2}{c_0^2}\,\frac{\tau}{t_0}\sqrt{\mid E_1\mid}
+\ldots\right)\, , & \text{if} & \frac{t_0}{\tau (T_1)}\gg \frac{c_{\infty}^2}{c_0^2}\sqrt{\mid E_1\mid}
\\
\\
\frac{\mid E_1\mid}{2\pi}\,\frac{\tau}{t_0^2}\,\frac{c_{\infty}^2}{c_0^2}\, , & \text{if} &
\frac{c_{\infty}^2}{c_0^2}\sqrt{\mid E_1\mid}\gg\frac{t_0}{\tau (T_1)}\gg\frac{c_{\infty}}{c_0}\sqrt{\mid
E_1\mid}
\\
\\
\frac{c_{\infty}}{c_0}\,\frac{\sqrt{\mid
E_1\mid}}{2\pi}\,\frac{1}{t_0}\left(1-\,\frac{1}{2}\,\frac{c_{\infty}^2-c_0^2}{c_{\infty}^2}\,
\frac{c_0}{c_{\infty}} \sqrt{\mid E_1\mid}\frac{t_0}{\tau} +\ldots\right)\, , & \text{if} &
\frac{c_{\infty}}{c_0}\sqrt{\mid E_1\mid}\gg\frac{t_0}{\tau (T_1)}
\end{array}
\right.
\end{equation}
\end{widetext}
Value $E_1$ can also be found with the help of the variational principle, minimizing the functional
\begin{equation}\nonumber
E=\frac{\int d^3x\, [(\nabla\chi )^2+(1+2\chi _{\text{cl}}(\bm{x}))\chi ^2]}{\int d^3x\,\bm{w}^2} \;\; ,\;\;\;
\chi=-\nabla\cdot\bm{w}
\end{equation}
The approximate value $E_1\approx$ $-$6.11 yields temperature $T_1\sim 0.39/t_0$ correcting the estimate
$T_{q0}\sim 0.27/t_0$  for the case $\tau=0$ in \eqref{f28}. As is seen from \eqref{f41}, the calculations of
temperature $T_1$ for absolute instability of the classical path are in agreement with the estimates in
\eqref{f28}--\eqref{f33}. The temperature of the thermal-quantum crossover $T_q$ cannot be lower than $T_1$.

\section{Discussion}

\par
Discussing common implications for quantum decay of a metastable liquid, first of all we should emphasize that
the relaxation time is an important parameter governing the nucleation process. The quantum nucleation rate
proves to be a monotone increasing function of relaxation time $\tau$, saturating in the limit of infinite
relaxation time $\tau =\infty$. Depending on a ratio of high frequency sound velocity to low frequency one
$c_{\infty}/c_0$, we can distinguish either two or three quantum nucleation regime. If $c_{\infty}/c_0\gtrsim
1$, we have the low frequency or high frequency regime depending on the relationship between the relaxation
time and the vicinity to the spinodal. In the case of strong inequality  $c_{\infty}/c_0\gg 1$ it becomes
possible to discern the crossover between the low and high frequency regimes as an independent regime
corresponding to the overdamped viscous quantum nucleation.
\par
Second, provided the relaxation time $\tau (T)$ as a function of temperature diverges for $T\rightarrow 0$,
the nucleation rate $W=W(T)$ can demonstrate a non-monotone temperature behavior with a minimum in the region
of the thermal-quantum crossover temperature $T_q$. Accordingly, if the nucleation rate is fixed under
experimental conditions, the observable supersaturation of the metastable phase will be maximum. The larger
the ratio $c_{\infty}/c_0$, the more prominent the relative magnitude of the effect.
\par
The third aspect of nucleation concerns the behavior the thermal-quantum crossover temperature as one
approaches the spinodal point where the sound velocity vanishes $c_0(\rho _c)=0$. Keeping in mind $\tau
^{-1}(T\rightarrow 0)\rightarrow 0$, one can see that the crossover temperature $T_q$ reduces in a linear
proportion to the product $c_{\infty}(\rho)c_0(\rho)$ vanishing at the spinodal point $\rho =\rho _c$ together
with the potential barrier. Thus the nucleation process in the region of metastability immediately adjacent to
the spinodal should be governed by the classical thermal activation.
\par
Let us compare liquid $^3$He and $^4$He. In Bose-liquid $^4$He the thermal-quantum crossover temperature is
not expected to exceed approximately 1K. For definiteness, we consider the temperature region below 0.6K when
the contribution from rotons into all phenomena becomes insignificant and excitations in liquid can be treated
as a purely phonon gas. Since the process of sound propagation in $^4$He is closely connected with the
propagation of phonons, the velocities of both low frequency and high frequency sound do not differ much from
the phase velocity $c$ of phonons. The deviation from the velocity of phonons is wholly due to the presence of
thermal excitations with normal density $\rho _n = 2\pi ^2T^4/(45\hbar ^3c^5)$. The difference in velocities
$c_{\infty}$ and $c_0$ can be found using, e.g., \cite{r20}
\begin{eqnarray*}
c_{\infty}-c_0 & = & c\,\frac{\rho _n(T)}{\rho}\left[\frac{3}{4}(A+1)^2\log\frac{1}{\gamma}\left(\frac{c}{2\pi
T}\right)^2\ \right.
\\
& -  & \left. \frac{(3A+1)^2}{4}-3A-2\right]
\end{eqnarray*}
Here $A=\partial\log c/\partial\log\rho$ is the Gr\"{u}neisen parameter and $\gamma$ is the coefficient of the
cubic term describing deviation of the phonon dispersion from the linear one. The relaxation time in the
system of phonons grows drastically at low temperatures
\begin{equation}\nonumber
\frac{1}{\tau}\sim (A+1)^4\frac{\hbar ^2}{\rho ^2c}\left(\frac{T}{2\pi\hbar c}\right) ^9
\end{equation}
Such large relaxation time at $T\rightarrow 0$ results in temperature $T_{\tau 1}$ always higher than the
thermal-quantum crossover temperature $T_q$. In the vicinity of the spinodal one has $T_{\tau 1}\propto
c^{8/9}$ and crossover temperature $T_q\propto c_0c_{\infty}\sim c^2$. Thus the regime of quantum nucleation
corresponds to the high frequency limit and the underbarrier growth of a nucleus occurs under collisionless
conditions, i.e., typical size of a nucleus is much less compared with the mean free path of excitations. The
latter favors the quantum nucleation.
\par
However, the temperature effect associated with the distinction in velocities $c_{\infty}$ and $c_0$ and also
with their temperature dependence, which becomes of the order of 1ppm below 0.1K, is not large. The relative
contribution into the effective action has the order of the magnitude of small parameter $\rho _n/\rho$. In
addition, though the low frequency and high frequency regimes of quantum nucleation differs significantly in
kind from the physical point of view, the quantitative difference between the regimes is again moderate since
the relative change of the sound velocity from the low to the high frequency limit is about the same small
ratio $(c_{\infty}-c_0)/c\sim\rho _n/\rho$. From the experimental point of view this can hardly be discerned
unless a rather precise measurement of the nucleation rate is employed. On the whole one should not expect any
noticeable temperature variations in the rate of quantum nucleation at low temperatures. Apparently, such
picture takes place in the experiments on nucleation of solids or cavities in superfluid $^4$He where the
temperature-independent nucleation rate is observed at temperatures below one or more hundred of mK.
\par
Let us turn now to normal liquid $^3$He. In Fermi liquid the nature of the low and high frequency sound modes,
associated with the various physical mechanisms, differs in a qualitative sense. Thus, unlike superfluid
$^4$He, sound velocities $c_0$ and $c_{\infty}$ are different even at zero temperature. The relative
difference amounts about 4\% at zero pressure. The relaxation time due to collisions between $^3$He
quasiparticle excitations varies with temperature as
\begin{equation}\label{f45}
\tau =\nu /T^2\; \;\;\; (\nu\approx2\cdot 10^{-12}\; s\cdot K^2)
\end{equation}
Generally speaking, the numerical coefficient depends on pressure or density of a liquid.
\par
Within the framework of the Fermi liquid theory the instability at spinodal $c_0(\rho _c)=0$ implies for the
Landau parameter $F_0=-1$. However, velocity $c_{\infty}$ of the high frequency collisionless mode may remain
finite since the condition of the thermodynamic stability $c_0^2>0$ cannot be applied for nonequilibrium
processes. Unlike the usual case of $F_0>0$, the existence of high frequency zero-sound mode depends on the
magnitude of the next Landau parameter $F_1$ responsible for the value of the effective mass and Fermi
velocity of quasiparticles. The solution of the dispersion equation for zero-sound mode \cite{r20}
\begin{equation}\nonumber
\frac{s}{2}\log\frac{s+1}{s-1}\,-1= \frac{1+F_1/3}{F_0(1+F_1/3)+F_1s^2}
\end{equation}
where $s=c_{\infty}/v_F$ is a ratio of the velocity of zero-sound wave to the Fermi velocity, remains real at
$F_0=-1$ provided $F_1>3/2$. To estimate $F_1$ at the spinodal, we use an extrapolation \cite{r21} of the
dependence of the effective mass on density into the negative pressure region from the fit of the data at
positive pressures
\begin{equation}\nonumber
(1+F_1/3)^{-1}=m/m^{\star}=[1.0166(1-5.138\rho)]^2
\end{equation}
Substituting spinodal density $\rho _c=0.054$ g/cm$^3$ as estimated in \cite{r21}, we have approximately
$F_1\approx 3$ and arrive at zero-sound velocity almost same as the Fermi one $s=1.006$ instead of $F_1=6.25$
and $s=3.6$ at zero pressure. On the whole, reduction of the effective mass prevails over reduction of the
density and should result even in increasing the Fermi velocity by approximately one-third as compared with
the case of zero pressure. Eventually, the only essential point  is that the high and low frequency sound
velocities are expected to have various limiting behavior at negative pressures and their ratio
$c_{\infty}/c_0$ enhances infinitely with approaching the spinodal.
\par
Let us introduce temperature $T_{\nu}$ at which the thermal energy of excitations is about of quantum
uncertainty in energy due to collisions between  excitations, i.e.,
\begin{equation}\nonumber
\tau (T_{\nu})=1/T_{\nu}
\end{equation}
In essence, this temperature determines the upper limit of applicability of the Fermi-liquid theory.
According to \eqref{f45} $T_{\nu}=\nu\approx$0.26 K and is well below the Fermi temperature. The next
speculation depends the dimensionless parameter
\begin{equation}\nonumber
\zeta =\nu\tau (T_{\tau 2})=\frac{\nu\sqrt{\lambda\rho}}{c_{\infty}^2}
\end{equation}
Comparing temperatures $T_{\nu}$ and $T_{\tau 2}$ defined in \eqref{f27}, we find the relationship
\begin{equation}\nonumber
T_{\tau 2}=T_{\nu}\zeta ^{-1/2}
\end{equation}
\par
At first, we consider the most diversified case $\zeta\ll 1$ when $T_{\tau 2}>T_{\nu}$. In accordance with
\eqref{f41} we can discern three regions in the behavior of the thermal-quantum crossover temperature as a
function of the vicinity to the spinodal point
\begin{equation}\nonumber
T_q\simeq\left\{
\begin{array}{lcc}
\frac{c_{\infty}}{c_0}\,\frac{\sqrt{\mid E_1\mid}}{2\pi}\,\frac{1}{t_0}\propto c_0 \, , & \text{if} & c_0<
c_{\infty}\zeta
\\
\\
\left(\frac{\mid E_1\mid}{2\pi}\,\frac{\nu}{t_0^2}\,\frac{c_{\infty}^2}{c_0^2}\right)^{1/3}\propto c_0^{2/3}\,
, & \text{if} &  c_{\infty}\zeta< c_0< c_{\infty}\zeta ^{1/4}
\\
\\
\frac{\sqrt{\mid E_1\mid}}{2\pi}\,\frac{1}{t_0}\propto c_0^2 \, , & \text{if} & c_{\infty}\zeta ^{1/4}<c_0
\end{array}
\right.
\end{equation}
Thus, depending on the relation between $\zeta$ and $c_0/c_{\infty}$, we can observe either one, or two, or
three types of nucleation in the quantum region as the temperature varies from zero one to the thermal-quantum
crossover. While $c_0/c_{\infty}<\zeta$ one has only the high-frequency collisionless type. Within
intermediate region $\zeta <c_0/c_{\infty}<\zeta ^{1/4}$ there is a crossover to the overdamped type at $T\sim
T_{\tau 1}$ and at $T\sim T_{\tau 2}$, if $c_0/c_{\infty}>\zeta ^{1/4}$, the nucleation process becomes a
low-frequency collisional one. Compared with the zero temperature value the effective action increases with
the temperature as $T^2$ due to enhancement of the collision frequency $\tau ^{-1}(T)$. Correspondingly, the
nucleation rate decreases. The relative effect at the thermal-quantum crossover temperature when the
nucleation rate is minimal is most prominent if $c_0/c_{\infty}>\zeta ^{1/4}$.
\par
The case when $\zeta >1$ is meagre. Here, since always $c_0<\zeta c_{\infty}$, in the quantum regime we can
have only the high-frequency collisionless regime with the next crossover to the thermal Arrhenius nucleation.
The temperature behavior of the effective action and nucleation rate demonstrates the same features as in the
above-considered case $\zeta \ll 1$ though not so well-marked.
\par
Concerning the numerical estimate of parameter $\zeta$, one should evaluate, first of all, parameter $\lambda$
related closely  with the spatial dispersion  of sound in the long-wavelength limit. Thus we need in the
reliable estimate of $\lambda$ near the spinodal point. As a rule, the spatial dispersion, associated with the
interatomic spacing $a$, becomes significant at wave vectors close to $k\sim 1/a$. To estimate the order of
the magnitude for $\lambda$, we put that nonlinear term in the sound dispersion due to $\lambda (\nabla\rho
)^2$ becomes comparable with the main linear term at $k\sim 1/a$. Since we suppose that $\lambda$ remains
finite at the spinodal, it is convenient to represent
\begin{equation}\nonumber
\lambda\rho \simeq c_{\infty}^2 a^2
\end{equation}
entailing $\zeta\simeq\nu a/c_{\infty}$. Involving that the Fermi momentum is also about $1/a$ and
$c_{\infty}$ is comparable with the Fermi velocity, we arrive at the dimensional estimate for parameter
$\zeta$
\begin{equation}\nonumber
\zeta \sim T_{\nu}/T_F
\end{equation}
So, in liquid $^3$He one may expect that $\zeta\lesssim 1$, entailing a noticeable minimum in the temperature
behavior of the nucleation rate in thermal-quantum crossover region. Provided the nucleation rate keeps
constant in experiment, this means that attainable deviation $\delta\rho =\rho -\rho _c$ and, correspondingly,
negative cavitation pressure $P$ pass through a minimum in the course of the crossover from the
thermal-to-quantum regime with the next almost temperature-independent behavior at sufficiently lower
temperatures. The similar temperature behavior of the supersaturation due to the effect of relaxation
processes on the quantum nucleation is observed in supersaturated $^3$He-$^4$He liquid mixtures \cite{r5}. As
it concerns the cavitation experiments in $^3$He \cite{r10}, in this sense the observation of a sharp increase
of the cavitation threshold at temperatures below about 60 mK can serve for some evidence of the incipient
thermal-quantum crossover.
\par
In conclusion, we have suggested a theory which, for the first time, involves the relaxation and high
frequency properties of a condensed medium into quantum decay of a metastable  liquid near the spinodal at low
temperatures. The model developed can be employed for clarifying the physical picture of the low temperature
cavitation in liquid $^3$He and $^4$He at negative pressures. The results obtained might thus be a helpful
guide for experiments on the quantum decay in a metastable condensed medium.

\begin{acknowledgments}
The work is supported in part by the Grants of RFBR Nos.~04-02-17363a, 02-02-16772a and School
No.~2032.2003.2.
\end{acknowledgments}

\end{document}